\documentclass[a4paper,fleqn,usenatbib]{mnras}

\usepackage[T1]{fontenc}
\usepackage{ae,aecompl}

\usepackage{graphicx}	% Including figure files
\usepackage{amsmath}	% Advanced maths commands
\usepackage{amssymb}	% Extra maths symbols
\usepackage{comment}

\title[HD~189733b Broadband Filter Spectrum]{A Transmission Spectrum Of HD~189733b From Multiple Broadband Filter Observations}

\author[D. H. Kasper et al.]{
David H. Kasper,$^{1}$\thanks{E-mail: dkasper@uwyo.edu (WYO)} 
Jackson L. Cole,$^{1,2}$ 
Cristilyn N. Gardner,$^{1,3}$ 
Bethany R. Garver,$^{1,4}$ \and
Kyla L. Jarka,$^{1,5}$ 
Aman Kar,$^{1}$ 
Aylin M. McGough,$^{1}$ 
David J. PeQueen,$^{1,6}$ \and
Daniel Ivan Rivera,$^{1,7}$ 
Hannah Jang-Condell,$^{1}$ 
Henry A. Kobulnicky,$^{1}$\and
Adam D. Myers,$^{1}$
and Daniel A. Dale$^{1}$
\\
$^{1}$Department of Physics \& Astronomy, University of Wyoming, Laramie, WY 82072, USA\\
$^{2}$Physics and Astronomy Department, Middle Tennessee State University, Murfreesboro, TN 37132, USA\\
$^{3}$Department of Physics, California State University San Bernardino, San Bernardino, CA 92407, USA\\
$^{4}$Physics Department, Seattle Pacific University, Seattle, WA 98119, USA\\
$^{5}$Physics Department, Colorado College, Colorado Springs, CO 80903, USA\\
$^{6}$Department of Physical Sciences, Embry-Riddle Aeronautical University, Daytona Beach, FL 32114, USA\\
$^{7}$Department of Astronomy, San Diego State University, San Diego, CA 92182, USA
}

\date{Accepted MNRAS 2018 December 6}

\pubyear{2018}

\begin{document}
\label{firstpage}
\pagerange{\pageref{firstpage}--\pageref{lastpage}}
\maketitle

\begin{abstract}
We present new multi-broadband transit photometry of HD~189733b obtained at the Wyoming Infrared Observatory. Using an ensemble of five-band Sloan filter observations across multiple transits we have created an ``ultra-low'' resolution transmission spectrum to discern the nature of the exoplanet atmosphere. The observations were taken over three transit events and total 108 $u'$, 120 $g'$, 120 $r'$, 110 $i'$, and 116 $z'$ images with an average exposure cadence of seven minutes for an entire series. The analysis was performed with a Markov-Chain Monte-Carlo method assisted by a Gaussian processes regression model. We find the apparent planet radius to increase from $0.154^{+0.00092}_{-0.00096}$ $R_{*}$ at $z'$-band to $0.157^{+0.00074}_{-0.00078}$ $R_{*}$ at $u'$-band. Whether this apparent radius implies an enhanced Rayleigh scattering or clear or grey planet atmosphere is highly dependent on stellar spot modeling assumptions, but our results are consistent with the literature for HD 189733b. This set of observations demonstrates the ability of our 2.3-m ground-based observatory to measure atmospheres of large exoplanets.
\end{abstract}


%%%%%%%%%%%%%%%%%%%%%%%%%%%%%%%%%%%%%%%%%%%%%%%%%%

%%%%%%%%%%%%%%%%% BODY OF PAPER %%%%%%%%%%%%%%%%%%

\section{Introduction}

%%general notes, always chronological references, make sure groups aren't left out
%a.general introduction to transit spectroscopy 'Transit spectroscopy has become a powerful tool for probing exoplanet characteristics'

From the first photometric confirmation of a transiting extrasolar planet, \citep{2000ApJ...529L..45C, 2000ApJ...529L..41H} the possibility of observing transit events spectrally to further characterize the exoplanet was discussed. The number of known transiting exoplanets has increased dramatically \citep[NASA Exoplanet Archive,][]{2013PASP..125..989A}, and transit observation techniques \citep{2001PASP..113.1428E} and analyses \citep{2002ApJ...580L.171M} have been developed. In particular, transmission spectroscopy has become a powerful tool for probing giant exoplanet atmospheres, as it reveals the absorption and scattering by the exoplanetary atmosphere on the host star's light during a transit \citep{2000ApJ...537..916S, 2001ApJ...553.1006B}. The infrared and optical signatures of the atomic, molecular, and particulate components have revealed the composition of dozens of giant exoplanet atmospheres to date \citep[recent e.g.][]{2016Natur.529...59S, 2018AJ...155..156T}. 

%b. lit a space based 'gold standard' but precious time and effectiveness is key

Space-based observations are the ``gold-standard'' for exoplanet transmission spectroscopy \citep{2002ApJ...568..377C, 2013MNRAS.432.2917P, 2014Natur.505...69K}. The spectrophotometric precision required to measure typical transmission effects is at the level of one part in $10^{4}$ or better --- a regime where space-based observations excel, especially at near-infrared wavelengths where Earth's atmospheric transmission can be poor and highly variable. Specialized observation techniques and heroic calibration efforts have permitted detections of exoplanet transmission features more recently to levels of few parts in $10^{5}$ \citep{2013ApJ...774...95D, 2016AJ...152..203L, 2016ApJ...832..202T}. Transmission spectroscopy of giant exoplanets has detected, for example, atomic species including \ion{Na}{I} in HD 209458b \citep{2002ApJ...568..377C}, a conspicuous lack of \ion{Na}{I} in WASP-12b \citep{2013MNRAS.436.2956S}, and molecular species including H$_{2}$ in HD 209458b \citep{2008AA...485..865L}, and H$_{2}$O in X0-1b \citep{2013ApJ...774...95D}. In addition, transmission spectroscopy has revealed continuum slopes that indicate Rayleigh scattering in exoplanet atmospheres \citep[e.g., HD~189733b;][]{2008AA...481L..83L} as well as atmospheres large enough to undergo hydrodynamic escape such as HD 209458b \citep{2003Natur.422..143V}. Sizeable samples of exoplanet transmission spectra demonstrate that their atmospheres range from clear to cloudy \citep{2015MNRAS.446.2428S, 2016Natur.529...59S, 2016ApJ...817L..16S, 2017MNRAS.464.4247W, 2018AJ...155..156T}. Clear atmospheres produce strong spectral features dictated by atmospheric composition and temperatures. Completely cloudy atmospheres yield featureless spectra. Most observed atmospheres lie somewhere between these extremes, with transmission spectra exhibiting discrete absorption features and a blue continuum slope. These spectra are best modeled as a mixture of optically thin molecular and atomic species as well as semi-transparent hazes and optically thick cloud decks. While the number of known transiting exoplanets has swelled into the thousands, the number with high-quality, broad-coverage, transmission spectra only approaches three dozen. Space-based measurements have been limited by the paucity of suitably instrumented observatories, motivating the need for ground-based efforts. 

%c. lib b ground based narrow band and med-low res difficult but time is more possible

Though ground-based observations suffer the disadvantages mentioned above,  their greater availability has produced significant results \citep{2008ApJ...673L..87R, 2010Natur.468..669B, 2017MNRAS.468.3907K}.
By simultaneously observing an ensemble of nearby field stars in addition to the host star, atmospheric effects can be removed to the precision necessary to recover the transit in detail. Typically ground-based spectroscopy employs single-slit differential spectrophotometry with a single comparison star \citep{2012MNRAS.426.1663S}, but multiobject spectrograph efforts have also been successful \citep[e.g.][]{2013ApJ...771..108B,2017AJ...154..242B}. Important comparison star considerations include non-variability, similarity in color, brightness, and angular proximity to the exoplanet host star such that simultaneous spectra can be recorded on the same instrument, allowing telluric variation to be removed from the target spectrum. The first ground-based observations successfully reproduced space-based results, namely the \ion{Na}{I} features, for example in HD 209458b, thereby demonstrating the capability of ground-based transmission spectroscopy \citep{2008ApJ...673L..87R}. Ground-based observations have since made original contributions, such as measurement of atmospheric circulation on HD 209458b \citep{2010Natur.465.1049S} and a continuum slope consistent with Rayleigh scattering in GJ 3470b \citep{2017AA...600A.138C}. Recently, many survey programs have been implemented to increase the number of high-quality broad-coverage transmission spectra, creating a statistically significant sample to understand atmospheric properties and their connection to planetary characteristics. \citep{2014AA...563A..41M, 2016ApJ...832..191N, 2017ApJ...834..151R, 2017AJ...154...95H}.  

%d. ground broadband
Multi-color broadband photometry is also a fruitful technique to investigate exoplanetary atmospheres  \citep{2012AA...538A..46D, 2017MNRAS.472.3871T}. This observational method principally measures the apparent variation of the planet radius as a function of wavelength, effectively creating an ``ultra-low'' resolution transmission spectrum. Some observational programs have employed dichroics to record a transit in multiple filters simultaneously, typically with large ($\geq 4$ m) telescopes \citep{2013AA...553A..26N, 2013AA...559A..32N, 2014MNRAS.437.1511B}. Other programs have utilized single near-ultraviolet filters where effects of Rayleigh scattering are most pronounced \citep{2016MNRAS.459..789T, 2017AN...338..773M}. Additionally some coordinated campaigns utilizing several (typically $\leq 1$ m) observatories have also produced multi-band transmission photometry \citep{2012AA...538A..46D, 2013MNRAS.436...2M, 2014AA...565A...7C, 2015ApJ...814..102D}. The highest precision observations (at a few parts in $10^{4}$) are achieved ultimately by maximizing photon counts through combinations of large collecting area \citep{2013AA...559A..32N} and/or simultaneous/repeated observations \citep{2013AA...553A..26N, 2013MNRAS.436...2M}. Multi-color broadband photometry is an efficient means to demonstrate the existence of a measurable exoplanet atmosphere and constrain its gross properties, while providing a prioritized target list for further study.

%f. in this work...

This paper presents new ground-based multi-broadband photometric transit observations of the extrasolar planet HD~189733b, a hot Jupiter well-studied with transmission spectroscopy \citep{2011MNRAS.416.1443S, 2012MNRAS.422..753G, 2012MNRAS.422.2477H, 2012AA...543L...4L, 2013MNRAS.432.2917P, 2015AA...577A..62W}. Our goal is to demonstrate the potential of an observational technique on a 2.3-m telescope through broadband filters to characterize exoplanet atmospheres. Dedicated access to such a telescope can yield, on its own, a high-quality ``ultra-low'' resolution transmission spectrum over just a few transits. HD~189733b, a $1.150 \pm 0.039$ M$_{\text{Jup}}$ exoplanet orbiting a favorably bright V$=7.648$ magnitude host \citep{2010MNRAS.403.1949K}, was chosen because of the favorable $2.2$ day period, transit duration of $1.8$ hours, and  planet-to-star radius ratio of $0.157$ \citep{2015MNRAS.450.3101B}. Additional characteristics of the exoplanet's orbit that informed our analysis were the $8.98$ ratio of orbital distance to stellar radius,  $85\fdg78$ inclination, and zero eccentricity \citep{2010MNRAS.408.1689S}. Section \ref{sec:observations} describes the observational procedure, Section \ref{sec:data_reduction} outlines the photometric reduction, Section \ref{sec:data_analysis} details the analysis employed to obtain the planet-to-star radius ratio as a function of wavelength, Section \ref{sec:star_spots} elaborates on our star spot corrections, and Section \ref{sec:discussion} discusses the success of the methodology, Section \ref{sec:conclusion} summarizes our conclusions.

\section{Observations at the Wyoming Infrared Observatory}
\label{sec:observations}
%nights observed, telescope, instrument
Observations of the HD~189733 system were taken on the UTC dates of 2016 August 2, 2017 June 22, and 2017 Aug 1 at the Wyoming Infrared Observatory (WIRO) 2.3-meter telescope in the Sloan $u'$, $g'$, $r'$, $i'$, and $z'$ filters. Table \ref{tab:obstable} provides a listing of exposures on each UTC date, with total exposures per filter, exposure length, the cadence (the time between successive exposures), and the airmass range. Images were acquired with the WIRO DoublePrime prime-focus imager \citep{2016PASP..128k5003F} which has a 0.58\arcsec\ pixel$^{-1}$ pixel scale on 4096$\times$4096 sky pixels resulting in a conveniently large 39\arcmin$\times$39\arcmin\ field of view. The instrument was used in full-field, four-amplifier mode, resulting in a 20 second read-out time.

%observation technique
Sequences of exposures alternating through the five filters were obtained continuously during each night. The telescope was purposely defocused such that the stellar point-spread-function was $\sim1\arcmin$ $\simeq$ 100 pixels in diameter to achieve exposure times that maximized photon counts while avoiding pixel saturation, averaging over interpixel variation noise, smoothing temporal sky variations, and reducing the effect of telescope tracking errors \citep{2010arXiv1001.2010W}. Exposure times in all filters were chosen on a nightly basis to achieve $\geq 10^{8}$ photons in each image for the target star, save for the $u'$-filter exposures on 2016 Aug 2 which obtained $\approx$ few$\times10^{7}$ photons per image for the target star. Care was given to ensure the target host's peak pixel counts remained well within the linear response regime of the detector. Exposure times over all nights ranged from 12 seconds in $r'$ to 320 seconds in $u'$. A drawback to this approach was the significance of scintillation noise in the uncertainty budget for all but the $u'$-band observations. For example, using the \citet{2015MNRAS.452.1707O} modified \citet{1967AJ.....72..747Y} scintillation approximation with the shortest exposures gives 0.62 mmags uncertainty at 1 airmass and 1.8 mmags uncertainty at 2 airmasses. This approach allowed us to obtain up to 17 in-transit exposures for each filter on a single night. Corrections to telescope tracking were performed by applying manual offsets between exposures on 2017 June 22 and on 2017 Aug 1 to maintain absolute image placement on the detector to within 10 pixels after the telescope was found to drift 40 pixels over the entire observation time on 2016 Aug 2. The target was not placed at the same pixel location between nights, with separations from night to night larger than the 50 pixel defocus radius.

%calibrations and time change
Image reductions involved first overscan-region subtraction, then bias subtraction, and finally flat-fielding with corresponding nightly sky flats in each filter.  Bias subtraction and overscan subtraction were applied to remove the bias pattern. Additionally, time transformations to Barycentric Julian Date in Barycentric Dynamical Time (BJD$_{\text{TDB}}$) from JD${_\text{UTC}}$ by mid-exposure time were completed with the \citet{2010PASP..122..935E} online converter. 

%notes on nights
Figure \ref{fig:weather_figure} shows the nightly raw target flux (left column) and sky background flux (right column) for each filter versus time. The time evolution of raw flux on the first night suggests somewhat clouded periods around BJD$_\text{TDB}$'s 0.86 and  beyond 0.89 on 2016 Aug 2. By comparison, the second and third night do not show evidence of clouded periods by raw flux time evolution.  The dramatic rise of the background flux rate at the end of the 2016 Aug 2 data is due to the approach of twilight. Likewise the dramatic fall of background flux rate on the 2017 Aug 1 data start is due to the end of twilight. On all nights the total background flux is $< 10^{-4}$ of the target flux, meaning the background contributes negligibly to the total noise. The decreased background flux after the first half of the 2017 Aug 1 observations, notable in $u'$, $g'$, and $r'$, is due to the moon setting.

\section{Photometric Data Reduction}
\label{sec:data_reduction}
%describe AIJ and comparison stars
We used the AstroImageJ (AIJ) package \citep{2013ascl.soft09001C, 2017AJ...153...77C}, which has been used extensively in exoplanet science in the KELT program \citep{2008AJ...135..907P,2012ApJ...761..123S, 2017Natur.546..514G, 2018AJ...155..100J} to perform photometry. AIJ performs differential aperture photometry to remove systematic differences between exposures and recover the target transit. Light curve generation was performed with AIJ on a single night, single filter basis, resulting in fifteen independent light curves for the HD~189733 system. We adopted a fixed aperture radius of 60 pixels and background annulus of 65 inner and 85 outer pixels that was kept constant between all nights and sources. These photometric parameters were chosen after experimentation over a range of possible values to maximize source signal and robustly measure the background. We selected four nearby field stars of similar apparent magnitude which are listed in Table \ref{tab:comps} with their brightness, color, and angular distance to the target star. We examined light curves of the comparison stars to verify their non-variability at the $\leq 2$ global mmag level over the course of the observations and to remove single images (less than five over the almost 600 total) that were clearly discrepant with the comparison ensemble trend.

%light-curve creation
Figure \ref{fig:multino_figure} left-hand side shows each filter's AIJ extracted photometric measurements from UTC 2016 Aug 2, UTC 2017 June 22, and UTC 2017 Aug 1 as color-coded from $u'$ in blue to $z'$ in black. Table \ref{table:photometry_table} presents all AIJ extracted relative photometry for all nights and filters. All measurements are plotted as folded around the best-fit, filter-dependent, mid-transit times and planet-to-star radius ratios as explained in Section \ref{sec:data_analysis}. The uncertainty budget for each photometric value is dominated by scintillation noise, approximated in the figure by the \citet{2015MNRAS.452.1707O} calculation.

\section{MCMC and GP Data Analysis}
\label{sec:data_analysis}
%introduce mcmc & gpr 
We analyzed the data through a Markov-Chain Monte Carlo (MCMC) method posterior retrieval using the {\tt emcee} \citep{2013PASP..125..306F} Python package to ultimately recover the planet's most probable radius as a function of wavelength. We utilized the {\tt batman} \citep{2015PASP..127.1161K} implementation of the \citet{2002ApJ...580L.171M} algorithm with a quadratic limb darkening law to generate transit models in the MCMC process. In generating comparison models, we accounted for photometric smearing in all data points by averaging {\tt batman} values along the exposure durations. We also utilized the Gaussian process (GP) regression modeler {\tt george} \citep{2015ITPAM..38..252A} to model and remove the noise-induced systematic trends in the data during the MCMC analysis. GP-assisted MCMC analysis has been used to detrend and analyze transmission and emission spectra of exoplanets with success \citep[recent e.g.][]{2017AJ...154..242B, 2018MNRAS.474..876K} and works by non-deterministically modeling the noise-induced correlation between data points in a simple functional way.

The reader is referred to \citet{2012MNRAS.419.2683G,2014MNRAS.445.3401G} for more information on the application of GP's to exoplanet transmission data, but here it suffices to give a short explanation of our choices following their methods. Given the variety of structure in the light curves, it was found insufficient to straight-forwardly detrend nightly filter series by low-order polynomial fits to auxiliary measurements (in particular airmass). Given the importance of temporal trends blended with changing total noise (itself likely dominated by scintillation noise and cloud interference) in the data set (Figure \ref{fig:multino_figure}: the temporal drift of relative target flux in the first night and last night, variability around mean model in all nights), we chose two kernels in our GP effort. To model the temporal trends (likely the result of uncorrected atmospheric transparency and instrumental drift) we utilize a squared-exponential kernel ($\Sigma$), shown in Equation \ref{eq:SQREXPeq} defined by a time-based length scale L$_{t}$ (longer than $1/2$ Transit duration) and an amplitude A$_{t}$ relating each light curve data point, $t_{n}$, to all other data points in the same light curve, $t_{m}$,

\begin{equation} \label{eq:SQREXPeq}
\Sigma_{n,m} = A_{t} \exp\left[-\left(\frac{t_{n}-t_{m}}{L_{t}}\right)^{2}\right],
\end{equation}

\noindent To model the white noise we utilize a white noise kernel defined with an amplitude A$_{wn}$ to be added to the diagonal elements of the covariance matrix based on the variance of the data (see below). These kernels were chosen after consideration of a handful of different kernel combinations that underperformed against this set.
%describe mcmc method

A simultaneous MCMC analysis of individual filter data sets was made. All orbital parameters necessary for {\tt batman} other than planet-to-star radius ratio and mid-transit time were fixed to previously published values, as listed in Table \ref{table:fixed_table}. These published values were set as fixed after ensuring parameter agreement between our data set and these published values. For each filter data set the MCMC process involved four free physical parameters: a single planet-to-star radius ratio and three mid-transit times. The priors for these four parameters were set broad and flat so as to let the data inform the parameters we were ultimately trying to constrain. The MCMC process was composed of 100 walkers on 100,000 steps each for each five independent filter series. Figure \ref{fig:multino_figure} contains the best fit MCMC solutions to each filter by night (left column) as well as a residual to the fit (right column) by filter with 500 randomly drawn corresponding MCMC sample models overplotted to give a sense for the goodness of the fit. Table \ref{table:midtransit_table} lists the by night, filter-averaged, mid-transit times in BJD$_{\text{TDB}}$ we fit which are in agreement with \citet{2015MNRAS.450.3101B}.

Quadratic limb darkening coefficients were assigned as fixed values for each filter from those quoted with the closest-matching stellar parameters in \citet{2011AA...529A..75C} (temperature of $T_\text{eff}=5000$~K and surface gravity of 4.5 log(cm~s$^{-2}$) as similar to calculated HD~189733 values from \citet{2017AJ...153..136S}, adopted metallicity of solar, and assumed microturbulent velocity of 0 km~s$^{-1}$). We also explored a small range around the adopted stellar parameters and concluded that the resulting small variations in limb darkening had a negligible effect on the final product as long as the coefficients all came from the same stellar parameter set.

In addition to the four physical parameters, the nine hyperparameters defining the GP kernels (three per night) were given flat and broad priors based on their definitions (see above). After early runs it became apparent that a singly-valued description of the data's nightly white noise amplitude was not always sufficient (as the residual r.m.s. to the best fit model changed appreciably on the first and second nights). For each filter's analysis of the 2016 Aug 02 data the white noise kernel was increased for the last 10 data points (as motivated by the all-filter flux drop in Figure \ref{fig:weather_figure}, top row left) by the ratio of the variance between those data and the pre-transit data. The white noise amplitude for each filter's in-transit data on 2017 June 22 was increased by the ratio of the variance between those data and the post-transit data. Although no cloudy period could be seen in the raw flux, the differing variance is noticeable in Figure \ref{fig:multino_figure}, middle row. These piece-wise changes to the white noise kernel represent only an improvement to the accuracy of the variance for those data points to be used in the diagonal of the covariance matrix. This piece-wise approach could be expanded into a fully realized noise envelope for the data, but such additional complexity was deemed unwarranted. The resulting white noise amplitudes were $\approx 10^{-6}$ for all August observations and $\approx 4\cdot 10^{-6}$ for the June observations. The resulting squared exponential kernel amplitudes were of similar order of magnitude with 2016 Aug observations $A_{t} \approx 10^{-6}, L_{t} \approx 0.14$ days, 2017 Aug observations $A_{t} \approx 5\cdot 10^{-6}, L_{t} \approx 0.026$ days and June observations resolving to a negligible kernel. These general estimates are useful to understand the importance of white noise in all nights/filters, the importance of the squared exponential kernel in the 2017 Aug observations (as can be seen in the data trends in the bottom-right panel of Figure \ref{fig:multino_figure}) and to a lesser extent the 2016 Aug observations, and the lack of squared exponential kernel importance for the June 22 data (which were dominated by white noise over the relatively shorter observation series). 

The lack of pre-transit baseline in both 2017 observations requires justification to support inclusion in the multi-night analysis. To this end each single-night, single-filter data set was independently run through a version of the above MCMC process adapted for single-night fitting. The resulting nightly planet-to-star radius ratios are shown in Figure \ref{fig:soloradius_figure}. The figure indicates that for the non-$u'$ filters the nights are in good agreement. Further there seems to be no systematic trend biasing one night through all five filters. The 2017 June 22 reported $u'$ filter ratio does appear to be significant in its offset from the other two nights, but due to the white noise envelope's relatively larger size (as seen in Figure \ref{fig:multino_figure}) on the in-transit data for that night (compared to the other nights) the 2017 June 22 contribution to the derived multi-night $u'$ ratio is weighted less (effectively the 2017 June 22 $u'$ ratio has a larger than quoted uncertainty).

\section{Star Spot Corrections}
\label{sec:star_spots}
HD~189733 is a chromospherically active star with many years worth of photometric measurements through the Automated Patrol Telescope photometer monitoring \citep{1999PASP..111..845H,2008AJ....135...68H}. As described in the literature on HD~189733 \citep{2011A&A...526A..12D,2011MNRAS.416.1443S,2013MNRAS.432.2917P,2014ApJ...791...55M} this necessitates corrections to the observed planet-to-star radius ratios for occulted and unocculted star spots. Given the time resolution, noise budget of our data, and uncorrelated residuals in time and color during transit we conclude there were no measured star spot crossings during our transit observations. Given the discrepancy between efforts utilized within the literature to correct HD~189733b transmission spectra for unocculted star spots, we endeavor to show agreement between our data set and those previous by recreating the \citet{2013MNRAS.432.2917P} and \citet{2014ApJ...791...55M} unocculted star spot corrections to compare our corrected photometry to their results (we follow \citet{2014ApJ...791...55M} Section 5.2 for the limiting case of ``all of the observed increase in the blue is due to unocculted star spots''). For both the Pont and McCullough corrections we utilize the aggregate star spot model, ignoring limb darkening as in \citet{2014ApJ...791...55M} Equation 1,
\begin{equation} \label{specslope}
\frac{R'_{p}}{R'_{s}} = \frac{R_{p}}{R_{s}} \frac{1}{\sqrt{1-\delta(1-F_{\rm spot}{\lambda}/F_{\rm phot}{\lambda})}},
\end{equation}
where $'$s denote the observed radius ratio, $\delta$ is the fraction of projected stellar surface covered with spots, $F_{\rm spot}{\lambda}$ and $F_{\rm phot}{\lambda}$ denote the star spot and stellar photosphere radiances, respectively. The first factor on the right is the radius ratio that would be observed in the absence of unocculted star spots. Despite knowledge that the activity level has been measured to vary by as much as $1-2\%$ \citet{2013MNRAS.432.2917P}, we are forced to approximate every night's correction by the same average spot corrections. This was due to the varied placement on the CCD for each source between nights prohibiting sufficient absolute characterization of the out-of-transit stellar flux between nights. To our advantage, the spacing between the observations was significantly larger than the $\approx$12 day rotation period of HD~189733 \citep{2008AJ....135...68H} and, thus, each of the three nights absolute flux was effectively independent.

With this approach, two corrections can be made to our data set to compare to the two literature sources based on simple differences in the spot coverage fraction and spot temperature. To recreate the average \citet{2013MNRAS.432.2917P} spot covering fraction and spot temperature we utilize a $\delta= 0.01$, 4000~K spot population. To recreate the \citet{2014ApJ...791...55M} limiting case spot covering fraction and spot temperature we utilize a $\delta= 0.043$, 4250~K spot population correction. \citet{2014ApJ...791...55M} was concerned with an unobserved additional spot population beyond the \citet{2013MNRAS.432.2917P} population assumed through observed stellar variability. Such a spot population which does not change with time could have a large impact on the observed transit depth, but would require inhomogeneous distribution across the star to be unobserved by planet occultations. \citet{2014ApJ...791...55M} point to polar regions as a potential physical location. 

To generate the spot and photospheric radiances we utilized PHOENIX atmosphere models \citep{2013A&A...553A...6H} of $\log(g) = 4.5$ and $[M/H] = 0.0$ (consistent with \citet{2017AJ...153..136S} and our earlier limb darkening assumptions). Table \ref{tab:results} first lists uncorrected planet-to-star radius ratios for our multi-night analysis of each filter followed by the resulting spot-corrected values. The top of Figure \ref{fig:corrections_figure} shows the uncorrected $R_{\text{p}}/R_{\text{s}}$ (in blue) with both our Pont spot covering fraction and spot temperature (in orange, with slight midpoint offset for clarity) and our McCullough limiting case spot covering fraction and spot temperature (in green) against wavelength of observation, where the filter bandpasses are represented by the horizontal error bars. Overplotted for convenience are the \citet{2013MNRAS.432.2917P} (in red) and \citet{2014ApJ...791...55M} limiting case (in purple) data series, both offset by -0.001 $R_{\text{p}}/R_{\text{s}}$ due to limb darkening correction differences between our data and the two literature sources.

For the Pont spot covering fraction and spot temperature points, the increasing exoplanet radius towards shorter wavelengths from $\approx0.1537$ to $\approx0.1565$  indicates the exoplanet appears bigger at bluer wavelengths, ostensibly the signature of an atmosphere. Given the expectation of a linear relation between exoplanet size and $\ln$(wavelength) we ran a 100-walker 10,000-step MCMC linear fit of our data to retrieve the most probable slope. Figure \ref{fig:mbfit_figure} shows the data points (black) with the quoted fit overplotted denoting the $\pm1\sigma$ (dark red-dashed) and $\pm2\sigma$ (light red-dashed) slope uncertainties, respectively. The retrieved slope is $\frac{\text{d}R_{\text{p}}/R_{\text{s}}}{\text{dln}\lambda} = -0.0029 \pm 0.0011 \ln(\text{nm})^{-1}$, which argues against a grey atmosphere in the observed range at the 99\% (2.6$\sigma$) level.

By following the general arguments and Equation 4 in \citet{2008AA...481L..83L}, we can apply,

\begin{equation} \label{specslope}
\alpha T = \frac{\mu g}{k}\frac{\text{d}R_{p}}{\text{dln}\lambda},
\end{equation}

\noindent to relate the slope $\frac{\text{d}R_{p}}{\text{dln}\lambda}$ in the observed spectral regime to a temperature ($T$) of the observed atmosphere where $\mu$ is the mean molecular weight of the atmospheric particles, $g$ is the local gravity, and $k$ is the Boltzmann constant. In calculating the temperature we elect to make the assumption that the $\frac{\text{d}R_{p}}{\text{dln}\lambda}$ slope is singly valued across the observed wavelengths. This decision is a simplification made despite knowledge of a break at 600 nm by more detailed, space-based, studies \citep{2008AA...481L..83L,2013MNRAS.432.2917P} because it is appropriate given the quality of the current data set. By multiplying the $\frac{\text{d}R_{\text{p}}/R_{\text{s}}}{\text{dln}\lambda}$ result by the stellar radius of $0.75 R_{\odot}$ \citep{2017AJ...153..136S}, we find an average slope of $\frac{\text{d}R_{p}}{\text{dln}\lambda}= -1513 \pm 574$ km from 320 nm to 1000 nm.  Assuming the mean mass of atmospheric particles is 2.3 times the mass of the proton, calculating the exoplanet surface gravity (21.9 $m/s^2$)\citep{2017AJ...153..136S}, and assuming Rayleigh scattering ($\alpha = -4$), we derive a temperature of $T= 2401 \pm 911$~K which agrees with the 2000 K 300--600 nm and is 1.2$\sigma$ from the 1300 K 600--1000 nm models adopted in \citet{2013MNRAS.432.2917P} as well as the $T= 2300 \pm 900$~K results from \citet{2015A&A...580A..84D} with precision more similar with our own. The arguments of \citet{2008AA...481L..83L} Section 4 can be applied to our result as well, implying Rayleigh scattering by well-mixed grains (haze) is preferred over molecular hydrogen as the Rayleigh scatterer. As in \citet{2008AA...481L..83L}, our preference is driven by the molecular hydrogen Rayleigh-scattering scenario requiring higher pressures that would lead to other abundant species (assuming solar-like abundances) signatures overcoming the Rayleigh signature.

{By contrast, a spot correction with a covering fraction of 4.3\% and spot temperature of 4250~K in Figure \ref{fig:corrections_figure} results in our planet-to-star radius ratio slope consistent with zero slope at 1.5-$\sigma$. This is in agreement with the \citet{2014ApJ...791...55M} limiting case analysis and the implication of a grey atmosphere for HD~189733b. This correction leads to a physical interpretation that for the atmosphere that is inconsistent with the conclusions of \citet{2013MNRAS.432.2917P} and our own conclusions following a $\delta= 0.01$, 4000~K spot population correction. The bottom of Figure \ref{fig:corrections_figure} shows a variety of potential unocculted spot corrections over a magnitude of potential coverage fractions. The intial 5.2 limiting case and final spot correction argued for by \citet{2014ApJ...791...55M} are also shown. Figure \ref{fig:corrections_figure} shows that for preferentially unocculted spot corrections $R_{\text{p}}/R_{\text{s}}$ is reduced for all filters with a greater relative effect in the bluer filters as coverage fraction increases. At the lowest coverage corrections, the exoplanet is required to have Rayleigh scattering hazes to account for the large spectral slope. \citet{2014ApJ...791...55M} argues for $\delta= 0.056$, 3700~K at which the stellar spot correction applied leaves the atmospheric interpretation as clear, without clouds or hazes. Eventually the proposed coverage fraction is large enough that the correction results in a grey atmosphere and beyond that the unphysically motivated situation of larger effective size in the redder optical wavelengths. 

The reduction of $R_{\text{p}}/R_{\text{s}}$ for all filters with any spot correction is particularly important in constraining potential over-interpretation in our by-night uncorrected $R_{\text{p}}/R_{\text{s}}$. As seen in Figure \ref{fig:soloradius_figure}, each night's filter ensemble can not be adjusted uniformly by a single spot correction to account for the low confidence discrepancy between the nights --- implying that an uncorrected noise source is the dominant source in the scatter seen. The low coverage fraction corrections ($\leq 6\%$), all of which are eminently reasonable given the limited understanding of the stellar spot population, highlight the complexity of disentangling stellar and planetary signals around active stars.

\section{Discussion}
\label{sec:discussion}

Our multiple, near-simultaneous, broadband photometric observations over three transits of HD~189733b were analyzed to successfully recover the exoplanet size with a precision necessary to retrieve atmospheric properties. These results demonstrate the capability of 2-m-class ground-based observations over just a few observed transits to explore exoplanet atmospheres. We show that for HD~189733b, interpretation of the observed exoplanet-to-star radius ratio is dominated by stellar spot assumptions inferred from previous literature observations. Outside of stellar activity concerns, a comprehensive list of the supplemental data required to perform this analysis included modestly constrained exoplanet orbital properties and modeled stellar limb darkening coefficients --- values that can be found in the literature for all confirmed exoplanets.

Inclusion of similarly bright comparison stars in the field was an important limiting factor in this analysis, as the $u'$-filter results can attest, and preconsideration of the best orientation/placement of the field will result in the best use of this methodology. We found that the $<60$ sec exposures suffered non-negligibly from additional r.m.s. variations about the best-fit likely due to unaveraged sky variations. We also found that the filter/observation sequencing described in Section \ref{sec:observations} led to smaller photometric residuals than a similar strategy that employed multiple exposures in each filter before switching to the next due to the more rapid cadence. Unsurprisingly, we found full transits better constraining then partials. Improvement in the ultimate measurements could be made in future efforts by measuring stellar psf's on the same pixels from night to night to do absolute calibration of the active star flux to $<1\%$ necessary to constrain the relative stellar activity.

\section{Conclusions}
\label{sec:conclusion}
We have obtained $>550$ photometric measurements of HD~189733 over the course of three HD~189733b transit events. The measurements were spread approximately equally over the five Sloan filters and enable a characterization of the exoplanet's atmosphere.  We find the existence of a measurable atmospheric response across the visible for HD~189733b is highly dependent on the unocculted star spot assumptions made. With a $\delta=0.01$, T=4000~K spot correction we measure the atmosphere of the exoplanet to have a spectral slope of $\frac{\text{d}R_{p}}{\text{dln}\lambda}= -1513 \pm 574$ km from 320 nm to 1000 nm. In contrast, with a spot correction of $\delta= 0.043$, 4250~K we find a spectral slope indicative of a grey atmosphere for the exoplanet.

We have shown the Wyoming Infrared Observatory is capable of the precision needed to attain exoplanet atmospheric characterization in transiting systems. In particular the 2.3-m observatory would be efficient at characterizing exoplanets hosted by less active stars.

\section*{Acknowledgements}
The Wyoming Infrared Observatory is owned and operated by the University of Wyoming.

This research has made use of the NASA Exoplanet Archive, which is operated by the California Institute of Technology, under contract with the National Aeronautics and Space Administration under the Exoplanet Exploration Program.

This research has made use of the SIMBAD database,
operated at CDS, Strasbourg, France.

This research was supported by The National Science Foundation through grant AST-1560461 (REU).

This research was supported by NASA XRP through grant NNX15AE23G.

The authors would like to thank the anonymous reviewer for their thoughtful suggestions as well as Thomas Beatty, Laura Kreidberg, and Neale Gibson for their fruitful discussions.

%%%%%%%%%%%%%%%%%%%%%%%%%%%%%%%%%%%%%%%%%%%%%%%%%%



\bibliographystyle{mnras}
\bibliography{2018bibliography} % if your bibtex file is called example.bib

\begin{thebibliography}{}
\makeatletter
\relax
\def\mn@urlcharsother{\let\do\@makeother \do\$\do\&\do\#\do\^\do\_\do\%\do\~}
\def\mn@doi{\begingroup\mn@urlcharsother \@ifnextchar [ {\mn@doi@}
  {\mn@doi@[]}}
\def\mn@doi@[#1]#2{\def\@tempa{#1}\ifx\@tempa\@empty \href
  {http://dx.doi.org/#2} {doi:#2}\else \href {http://dx.doi.org/#2} {#1}\fi
  \endgroup}
\def\mn@eprint#1#2{\mn@eprint@#1:#2::\@nil}
\def\mn@eprint@arXiv#1{\href {http://arxiv.org/abs/#1} {{\tt arXiv:#1}}}
\def\mn@eprint@dblp#1{\href {http://dblp.uni-trier.de/rec/bibtex/#1.xml}
  {dblp:#1}}
\def\mn@eprint@#1:#2:#3:#4\@nil{\def\@tempa {#1}\def\@tempb {#2}\def\@tempc
  {#3}\ifx \@tempc \@empty \let \@tempc \@tempb \let \@tempb \@tempa \fi \ifx
  \@tempb \@empty \def\@tempb {arXiv}\fi \@ifundefined
  {mn@eprint@\@tempb}{\@tempb:\@tempc}{\expandafter \expandafter \csname
  mn@eprint@\@tempb\endcsname \expandafter{\@tempc}}}

\bibitem[\protect\citeauthoryear{{Akeson} et~al.,}{{Akeson}
  et~al.}{2013}]{2013PASP..125..989A}
{Akeson} R.~L.,  et~al., 2013, \mn@doi [\pasp] {10.1086/672273}, \href
  {http://adsabs.harvard.edu/abs/2013PASP..125..989A} {125, 989}

\bibitem[\protect\citeauthoryear{{Ambikasaran}, {Foreman-Mackey}, {Greengard},
  {Hogg}  \& {O'Neil}}{{Ambikasaran} et~al.}{2015}]{2015ITPAM..38..252A}
{Ambikasaran} S.,  {Foreman-Mackey} D.,  {Greengard} L.,  {Hogg} D.~W.,
  {O'Neil} M.,  2015, \mn@doi [IEEE Transactions on Pattern Analysis and
  Machine Intelligence] {10.1109/TPAMI.2015.2448083}, \href
  {http://adsabs.harvard.edu/abs/2015ITPAM..38..252A} {38}

\bibitem[\protect\citeauthoryear{{Baluev} et~al.,}{{Baluev}
  et~al.}{2015}]{2015MNRAS.450.3101B}
{Baluev} R.~V.,  et~al., 2015, \mn@doi [\mnras] {10.1093/mnras/stv788}, \href
  {http://adsabs.harvard.edu/abs/2015MNRAS.450.3101B} {450, 3101}

\bibitem[\protect\citeauthoryear{{Bean}, {Miller-Ricci Kempton}  \&
  {Homeier}}{{Bean} et~al.}{2010}]{2010Natur.468..669B}
{Bean} J.~L.,  {Miller-Ricci Kempton} E.,   {Homeier} D.,  2010, \mn@doi [\nat]
  {10.1038/nature09596}, \href
  {http://adsabs.harvard.edu/abs/2010Natur.468..669B} {468, 669}

\bibitem[\protect\citeauthoryear{{Bean}, {D{\'e}sert}, {Seifahrt},
  {Madhusudhan}, {Chilingarian}, {Homeier}  \& {Szentgyorgyi}}{{Bean}
  et~al.}{2013}]{2013ApJ...771..108B}
{Bean} J.~L.,  {D{\'e}sert} J.-M.,  {Seifahrt} A.,  {Madhusudhan} N.,
  {Chilingarian} I.,  {Homeier} D.,   {Szentgyorgyi} A.,  2013, \mn@doi [\apj]
  {10.1088/0004-637X/771/2/108}, \href
  {http://adsabs.harvard.edu/abs/2013ApJ...771..108B} {771, 108}

\bibitem[\protect\citeauthoryear{{Beatty}, {Madhusudhan}, {Pogge}, {Chung},
  {Bierlya}, {Gaudi}  \& {Latham}}{{Beatty} et~al.}{2017}]{2017AJ...154..242B}
{Beatty} T.~G.,  {Madhusudhan} N.,  {Pogge} R.,  {Chung} S.~M.,  {Bierlya} A.,
  {Gaudi} B.~S.,   {Latham} D.~W.,  2017, \mn@doi [\aj]
  {10.3847/1538-3881/aa94cf}, \href
  {http://adsabs.harvard.edu/abs/2017AJ...154..242B} {154, 242}

\bibitem[\protect\citeauthoryear{{Bento} et~al.,}{{Bento}
  et~al.}{2014}]{2014MNRAS.437.1511B}
{Bento} J.,  et~al., 2014, \mn@doi [\mnras] {10.1093/mnras/stt1979}, \href
  {http://adsabs.harvard.edu/abs/2014MNRAS.437.1511B} {437, 1511}

\bibitem[\protect\citeauthoryear{{Brown}}{{Brown}}{2001}]{2001ApJ...553.1006B}
{Brown} T.~M.,  2001, \mn@doi [\apj] {10.1086/320950}, \href
  {http://adsabs.harvard.edu/abs/2001ApJ...553.1006B} {553, 1006}

\bibitem[\protect\citeauthoryear{{C{\'a}ceres} et~al.,}{{C{\'a}ceres}
  et~al.}{2014}]{2014AA...565A...7C}
{C{\'a}ceres} C.,  et~al., 2014, \mn@doi [\aap] {10.1051/0004-6361/201321087},
  \href {http://adsabs.harvard.edu/abs/2014A\%26A...565A...7C} {565, A7}

\bibitem[\protect\citeauthoryear{{Charbonneau}, {Brown}, {Latham}  \&
  {Mayor}}{{Charbonneau} et~al.}{2000}]{2000ApJ...529L..45C}
{Charbonneau} D.,  {Brown} T.~M.,  {Latham} D.~W.,   {Mayor} M.,  2000, \mn@doi
  [\apjl] {10.1086/312457}, \href
  {http://adsabs.harvard.edu/abs/2000ApJ...529L..45C} {529, L45}

\bibitem[\protect\citeauthoryear{{Charbonneau}, {Brown}, {Noyes}  \&
  {Gilliland}}{{Charbonneau} et~al.}{2002}]{2002ApJ...568..377C}
{Charbonneau} D.,  {Brown} T.~M.,  {Noyes} R.~W.,   {Gilliland} R.~L.,  2002,
  \mn@doi [\apj] {10.1086/338770}, \href
  {http://adsabs.harvard.edu/abs/2002ApJ...568..377C} {568, 377}

\bibitem[\protect\citeauthoryear{{Chen}, {Guenther}, {Pall{\'e}}, {Nortmann},
  {Nowak}, {Kunz}, {Parviainen}  \& {Murgas}}{{Chen}
  et~al.}{2017}]{2017AA...600A.138C}
{Chen} G.,  {Guenther} E.~W.,  {Pall{\'e}} E.,  {Nortmann} L.,  {Nowak} G.,
  {Kunz} S.,  {Parviainen} H.,   {Murgas} F.,  2017, \mn@doi [\aap]
  {10.1051/0004-6361/201630228}, \href
  {http://adsabs.harvard.edu/abs/2017A\%26A...600A.138C} {600, A138}

\bibitem[\protect\citeauthoryear{{Claret} \& {Bloemen}}{{Claret} \&
  {Bloemen}}{2011}]{2011AA...529A..75C}
{Claret} A.,  {Bloemen} S.,  2011, \mn@doi [\aap]
  {10.1051/0004-6361/201116451}, \href
  {http://adsabs.harvard.edu/abs/2011A\%26A...529A..75C} {529, A75}

\bibitem[\protect\citeauthoryear{{Collins} \& {Kielkopf}}{{Collins} \&
  {Kielkopf}}{2013}]{2013ascl.soft09001C}
{Collins} K.,  {Kielkopf} J.,  2013, {AstroImageJ: ImageJ for Astronomy},
  Astrophysics Source Code Library (\mn@eprint {ascl} {1309.001})

\bibitem[\protect\citeauthoryear{{Collins}, {Kielkopf}, {Stassun}  \&
  {Hessman}}{{Collins} et~al.}{2017}]{2017AJ...153...77C}
{Collins} K.~A.,  {Kielkopf} J.~F.,  {Stassun} K.~G.,   {Hessman} F.~V.,  2017,
  \mn@doi [\aj] {10.3847/1538-3881/153/2/77}, \href
  {http://adsabs.harvard.edu/abs/2017AJ...153...77C} {153, 77}

\bibitem[\protect\citeauthoryear{{Deming} et~al.,}{{Deming}
  et~al.}{2013}]{2013ApJ...774...95D}
{Deming} D.,  et~al., 2013, \mn@doi [\apj] {10.1088/0004-637X/774/2/95}, \href
  {http://adsabs.harvard.edu/abs/2013ApJ...774...95D} {774, 95}

\bibitem[\protect\citeauthoryear{{D{\'e}sert} et~al.,}{{D{\'e}sert}
  et~al.}{2011}]{2011A&A...526A..12D}
{D{\'e}sert} J.-M.,  et~al., 2011, \mn@doi [\aap]
  {10.1051/0004-6361/200913093}, \href
  {http://adsabs.harvard.edu/abs/2011A%26A...526A..12D} {526, A12}

\bibitem[\protect\citeauthoryear{{Di Gloria}, {Snellen}  \& {Albrecht}}{{Di
  Gloria} et~al.}{2015}]{2015A&A...580A..84D}
{Di Gloria} E.,  {Snellen} I.~A.~G.,   {Albrecht} S.,  2015, \mn@doi [\aap]
  {10.1051/0004-6361/201526218}, \href
  {http://adsabs.harvard.edu/abs/2015A%26A...580A..84D} {580, A84}

\bibitem[\protect\citeauthoryear{{Dragomir}, {Benneke}, {Pearson},
  {Crossfield}, {Eastman}, {Barman}  \& {Biddle}}{{Dragomir}
  et~al.}{2015}]{2015ApJ...814..102D}
{Dragomir} D.,  {Benneke} B.,  {Pearson} K.~A.,  {Crossfield} I.~J.~M.,
  {Eastman} J.,  {Barman} T.,   {Biddle} L.~I.,  2015, \mn@doi [\apj]
  {10.1088/0004-637X/814/2/102}, \href
  {http://adsabs.harvard.edu/abs/2015ApJ...814..102D} {814, 102}

\bibitem[\protect\citeauthoryear{{Eastman}, {Siverd}  \& {Gaudi}}{{Eastman}
  et~al.}{2010}]{2010PASP..122..935E}
{Eastman} J.,  {Siverd} R.,   {Gaudi} B.~S.,  2010, \mn@doi [\pasp]
  {10.1086/655938}, \href {http://adsabs.harvard.edu/abs/2010PASP..122..935E}
  {122, 935}

\bibitem[\protect\citeauthoryear{{Everett} \& {Howell}}{{Everett} \&
  {Howell}}{2001}]{2001PASP..113.1428E}
{Everett} M.~E.,  {Howell} S.~B.,  2001, \mn@doi [\pasp] {10.1086/323387},
  \href {http://adsabs.harvard.edu/abs/2001PASP..113.1428E} {113, 1428}

\bibitem[\protect\citeauthoryear{{Findlay}, {Kobulnicky}, {Weger}, {Bucher},
  {Perry}, {Myers}, {Pierce}  \& {Vogel}}{{Findlay}
  et~al.}{2016}]{2016PASP..128k5003F}
{Findlay} J.~R.,  {Kobulnicky} H.~A.,  {Weger} J.~S.,  {Bucher} G.~A.,  {Perry}
  M.~C.,  {Myers} A.~D.,  {Pierce} M.~J.,   {Vogel} C.,  2016, \mn@doi [\pasp]
  {10.1088/1538-3873/128/969/115003}, \href
  {http://adsabs.harvard.edu/abs/2016PASP..128k5003F} {128, 115003}

\bibitem[\protect\citeauthoryear{{Foreman-Mackey}, {Hogg}, {Lang}  \&
  {Goodman}}{{Foreman-Mackey} et~al.}{2013}]{2013PASP..125..306F}
{Foreman-Mackey} D.,  {Hogg} D.~W.,  {Lang} D.,   {Goodman} J.,  2013, \mn@doi
  [\pasp] {10.1086/670067}, \href
  {http://adsabs.harvard.edu/abs/2013PASP..125..306F} {125, 306}

\bibitem[\protect\citeauthoryear{{Gaudi} et~al.,}{{Gaudi}
  et~al.}{2017}]{2017Natur.546..514G}
{Gaudi} B.~S.,  et~al., 2017, \mn@doi [\nat] {10.1038/nature22392}, \href
  {http://adsabs.harvard.edu/abs/2017Natur.546..514G} {546, 514}

\bibitem[\protect\citeauthoryear{{Gibson}}{{Gibson}}{2014}]{2014MNRAS.445.3401G}
{Gibson} N.~P.,  2014, \mn@doi [\mnras] {10.1093/mnras/stu1975}, \href
  {http://adsabs.harvard.edu/abs/2014MNRAS.445.3401G} {445, 3401}

\bibitem[\protect\citeauthoryear{{Gibson}, {Aigrain}, {Roberts}, {Evans},
  {Osborne}  \& {Pont}}{{Gibson} et~al.}{2012a}]{2012MNRAS.419.2683G}
{Gibson} N.~P.,  {Aigrain} S.,  {Roberts} S.,  {Evans} T.~M.,  {Osborne} M.,
  {Pont} F.,  2012a, \mn@doi [\mnras] {10.1111/j.1365-2966.2011.19915.x}, \href
  {http://adsabs.harvard.edu/abs/2012MNRAS.419.2683G} {419, 2683}

\bibitem[\protect\citeauthoryear{{Gibson} et~al.,}{{Gibson}
  et~al.}{2012b}]{2012MNRAS.422..753G}
{Gibson} N.~P.,  et~al., 2012b, \mn@doi [\mnras]
  {10.1111/j.1365-2966.2012.20655.x}, \href
  {http://adsabs.harvard.edu/abs/2012MNRAS.422..753G} {422, 753}

\bibitem[\protect\citeauthoryear{{Henry}}{{Henry}}{1999}]{1999PASP..111..845H}
{Henry} G.~W.,  1999, \mn@doi [\pasp] {10.1086/316388}, \href
  {http://adsabs.harvard.edu/abs/1999PASP..111..845H} {111, 845}

\bibitem[\protect\citeauthoryear{{Henry} \& {Winn}}{{Henry} \&
  {Winn}}{2008}]{2008AJ....135...68H}
{Henry} G.~W.,  {Winn} J.~N.,  2008, \mn@doi [\aj]
  {10.1088/0004-6256/135/1/68}, \href
  {http://adsabs.harvard.edu/abs/2008AJ....135...68H} {135, 68}

\bibitem[\protect\citeauthoryear{{Henry}, {Marcy}, {Butler}  \& {Vogt}}{{Henry}
  et~al.}{2000}]{2000ApJ...529L..41H}
{Henry} G.~W.,  {Marcy} G.~W.,  {Butler} R.~P.,   {Vogt} S.~S.,  2000, \mn@doi
  [\apjl] {10.1086/312458}, \href
  {http://adsabs.harvard.edu/abs/2000ApJ...529L..41H} {529, L41}

\bibitem[\protect\citeauthoryear{{H{\o}g} et~al.,}{{H{\o}g}
  et~al.}{2000}]{2000AA...355L..27H}
{H{\o}g} E.,  et~al., 2000, \aap, \href
  {http://adsabs.harvard.edu/abs/2000A\%26A...355L..27H} {355, L27}

\bibitem[\protect\citeauthoryear{{Huitson}, {Sing}, {Vidal-Madjar},
  {Ballester}, {Lecavelier des Etangs}, {D{\'e}sert}  \& {Pont}}{{Huitson}
  et~al.}{2012}]{2012MNRAS.422.2477H}
{Huitson} C.~M.,  {Sing} D.~K.,  {Vidal-Madjar} A.,  {Ballester} G.~E.,
  {Lecavelier des Etangs} A.,  {D{\'e}sert} J.-M.,   {Pont} F.,  2012, \mn@doi
  [\mnras] {10.1111/j.1365-2966.2012.20805.x}, \href
  {http://adsabs.harvard.edu/abs/2012MNRAS.422.2477H} {422, 2477}

\bibitem[\protect\citeauthoryear{{Huitson}, {D{\'e}sert}, {Bean}, {Fortney},
  {Stevenson}  \& {Bergmann}}{{Huitson} et~al.}{2017}]{2017AJ...154...95H}
{Huitson} C.~M.,  {D{\'e}sert} J.-M.,  {Bean} J.~L.,  {Fortney} J.~J.,
  {Stevenson} K.~B.,   {Bergmann} M.,  2017, \mn@doi [\aj]
  {10.3847/1538-3881/aa7f72}, \href
  {http://adsabs.harvard.edu/abs/2017AJ...154...95H} {154, 95}

\bibitem[\protect\citeauthoryear{{Husser}, {Wende-von Berg}, {Dreizler},
  {Homeier}, {Reiners}, {Barman}  \& {Hauschildt}}{{Husser}
  et~al.}{2013}]{2013A&A...553A...6H}
{Husser} T.-O.,  {Wende-von Berg} S.,  {Dreizler} S.,  {Homeier} D.,  {Reiners}
  A.,  {Barman} T.,   {Hauschildt} P.~H.,  2013, \mn@doi [\aap]
  {10.1051/0004-6361/201219058}, \href
  {http://adsabs.harvard.edu/abs/2013A%26A...553A...6H} {553, A6}

\bibitem[\protect\citeauthoryear{{Johnson} et~al.,}{{Johnson}
  et~al.}{2018}]{2018AJ...155..100J}
{Johnson} M.~C.,  et~al., 2018, \mn@doi [\aj] {10.3847/1538-3881/aaa5af}, \href
  {http://adsabs.harvard.edu/abs/2018AJ...155..100J} {155, 100}

\bibitem[\protect\citeauthoryear{{Kirk}, {Wheatley}, {Louden}, {Doyle},
  {Skillen}, {McCormac}, {Irwin}  \& {Karjalainen}}{{Kirk}
  et~al.}{2017}]{2017MNRAS.468.3907K}
{Kirk} J.,  {Wheatley} P.~J.,  {Louden} T.,  {Doyle} A.~P.,  {Skillen} I.,
  {McCormac} J.,  {Irwin} P.~G.~J.,   {Karjalainen} R.,  2017, \mn@doi [\mnras]
  {10.1093/mnras/stx752}, \href
  {http://adsabs.harvard.edu/abs/2017MNRAS.468.3907K} {468, 3907}

\bibitem[\protect\citeauthoryear{{Kirk}, {Wheatley}, {Louden}, {Skillen},
  {King}, {McCormac}  \& {Irwin}}{{Kirk} et~al.}{2018}]{2018MNRAS.474..876K}
{Kirk} J.,  {Wheatley} P.~J.,  {Louden} T.,  {Skillen} I.,  {King} G.~W.,
  {McCormac} J.,   {Irwin} P.~G.~J.,  2018, \mn@doi [\mnras]
  {10.1093/mnras/stx2826}, \href
  {http://adsabs.harvard.edu/abs/2018MNRAS.474..876K} {474, 876}

\bibitem[\protect\citeauthoryear{{Koen}, {Kilkenny}, {van Wyk}  \&
  {Marang}}{{Koen} et~al.}{2010}]{2010MNRAS.403.1949K}
{Koen} C.,  {Kilkenny} D.,  {van Wyk} F.,   {Marang} F.,  2010, \mn@doi
  [\mnras] {10.1111/j.1365-2966.2009.16182.x}, \href
  {http://adsabs.harvard.edu/abs/2010MNRAS.403.1949K} {403, 1949}

\bibitem[\protect\citeauthoryear{{Kreidberg}}{{Kreidberg}}{2015}]{2015PASP..127.1161K}
{Kreidberg} L.,  2015, \mn@doi [\pasp] {10.1086/683602}, \href
  {http://adsabs.harvard.edu/abs/2015PASP..127.1161K} {127, 1161}

\bibitem[\protect\citeauthoryear{{Kreidberg} et~al.,}{{Kreidberg}
  et~al.}{2014}]{2014Natur.505...69K}
{Kreidberg} L.,  et~al., 2014, \mn@doi [\nat] {10.1038/nature12888}, \href
  {http://adsabs.harvard.edu/abs/2014Natur.505...69K} {505, 69}

\bibitem[\protect\citeauthoryear{{Lecavelier Des Etangs}, {Pont},
  {Vidal-Madjar}  \& {Sing}}{{Lecavelier Des Etangs}
  et~al.}{2008a}]{2008AA...481L..83L}
{Lecavelier Des Etangs} A.,  {Pont} F.,  {Vidal-Madjar} A.,   {Sing} D.,
  2008a, \mn@doi [\aap] {10.1051/0004-6361:200809388}, \href
  {http://adsabs.harvard.edu/abs/2008A\%26A...481L..83L} {481, L83}

\bibitem[\protect\citeauthoryear{{Lecavelier Des Etangs}, {Vidal-Madjar},
  {D{\'e}sert}  \& {Sing}}{{Lecavelier Des Etangs}
  et~al.}{2008b}]{2008AA...485..865L}
{Lecavelier Des Etangs} A.,  {Vidal-Madjar} A.,  {D{\'e}sert} J.-M.,   {Sing}
  D.,  2008b, \mn@doi [\aap] {10.1051/0004-6361:200809704}, \href
  {http://adsabs.harvard.edu/abs/2008A\%26A...485..865L} {485, 865}

\bibitem[\protect\citeauthoryear{{Lecavelier des Etangs} et~al.,}{{Lecavelier
  des Etangs} et~al.}{2012}]{2012AA...543L...4L}
{Lecavelier des Etangs} A.,  et~al., 2012, \mn@doi [\aap]
  {10.1051/0004-6361/201219363}, \href
  {http://adsabs.harvard.edu/abs/2012A\%26A...543L...4L} {543, L4}

\bibitem[\protect\citeauthoryear{{Line} et~al.,}{{Line}
  et~al.}{2016}]{2016AJ...152..203L}
{Line} M.~R.,  et~al., 2016, \mn@doi [\aj] {10.3847/0004-6256/152/6/203}, \href
  {http://adsabs.harvard.edu/abs/2016AJ...152..203L} {152, 203}

\bibitem[\protect\citeauthoryear{{Mallonn} \& {Wakeford}}{{Mallonn} \&
  {Wakeford}}{2017}]{2017AN...338..773M}
{Mallonn} M.,  {Wakeford} H.~R.,  2017, \mn@doi [Astronomische Nachrichten]
  {10.1002/asna.201713376}, \href
  {http://adsabs.harvard.edu/abs/2017AN...338..773M} {338, 773}

\bibitem[\protect\citeauthoryear{{Mancini} et~al.,}{{Mancini}
  et~al.}{2013}]{2013MNRAS.436...2M}
{Mancini} L.,  et~al., 2013, \mn@doi [\mnras] {10.1093/mnras/stt1394}, \href
  {http://adsabs.harvard.edu/abs/2013MNRAS.436...2M} {436, 2}

\bibitem[\protect\citeauthoryear{{Mandel} \& {Agol}}{{Mandel} \&
  {Agol}}{2002}]{2002ApJ...580L.171M}
{Mandel} K.,  {Agol} E.,  2002, \mn@doi [\apjl] {10.1086/345520}, \href
  {http://adsabs.harvard.edu/abs/2002ApJ...580L.171M} {580, L171}

\bibitem[\protect\citeauthoryear{{McCullough}, {Crouzet}, {Deming}  \&
  {Madhusudhan}}{{McCullough} et~al.}{2014}]{2014ApJ...791...55M}
{McCullough} P.~R.,  {Crouzet} N.,  {Deming} D.,   {Madhusudhan} N.,  2014,
  \mn@doi [\apj] {10.1088/0004-637X/791/1/55}, \href
  {http://adsabs.harvard.edu/abs/2014ApJ...791...55M} {791, 55}

\bibitem[\protect\citeauthoryear{{Murgas}, {Pall{\'e}}, {Zapatero Osorio},
  {Nortmann}, {Hoyer}  \& {Cabrera-Lavers}}{{Murgas}
  et~al.}{2014}]{2014AA...563A..41M}
{Murgas} F.,  {Pall{\'e}} E.,  {Zapatero Osorio} M.~R.,  {Nortmann} L.,
  {Hoyer} S.,   {Cabrera-Lavers} A.,  2014, \mn@doi [\aap]
  {10.1051/0004-6361/201322374}, \href
  {http://adsabs.harvard.edu/abs/2014A\%26A...563A..41M} {563, A41}

\bibitem[\protect\citeauthoryear{{Nascimbeni}, {Piotto}, {Pagano},
  {Scandariato}, {Sani}  \& {Fumana}}{{Nascimbeni}
  et~al.}{2013}]{2013AA...559A..32N}
{Nascimbeni} V.,  {Piotto} G.,  {Pagano} I.,  {Scandariato} G.,  {Sani} E.,
  {Fumana} M.,  2013, \mn@doi [\aap] {10.1051/0004-6361/201321971}, \href
  {http://adsabs.harvard.edu/abs/2013A\%26A...559A..32N} {559, A32}

\bibitem[\protect\citeauthoryear{{Nikolov}, {Chen}, {Fortney}, {Mancini},
  {Southworth}, {van Boekel}  \& {Henning}}{{Nikolov}
  et~al.}{2013}]{2013AA...553A..26N}
{Nikolov} N.,  {Chen} G.,  {Fortney} J.~J.,  {Mancini} L.,  {Southworth} J.,
  {van Boekel} R.,   {Henning} T.,  2013, \mn@doi [\aap]
  {10.1051/0004-6361/201321084}, \href
  {http://adsabs.harvard.edu/abs/2013A\%26A...553A..26N} {553, A26}

\bibitem[\protect\citeauthoryear{{Nikolov}, {Sing}, {Gibson}, {Fortney},
  {Evans}, {Barstow}, {Kataria}  \& {Wilson}}{{Nikolov}
  et~al.}{2016}]{2016ApJ...832..191N}
{Nikolov} N.,  {Sing} D.~K.,  {Gibson} N.~P.,  {Fortney} J.~J.,  {Evans} T.~M.,
   {Barstow} J.~K.,  {Kataria} T.,   {Wilson} P.~A.,  2016, \mn@doi [\apj]
  {10.3847/0004-637X/832/2/191}, \href
  {http://adsabs.harvard.edu/abs/2016ApJ...832..191N} {832, 191}

\bibitem[\protect\citeauthoryear{{Osborn}, {F{\"o}hring}, {Dhillon}  \&
  {Wilson}}{{Osborn} et~al.}{2015}]{2015MNRAS.452.1707O}
{Osborn} J.,  {F{\"o}hring} D.,  {Dhillon} V.~S.,   {Wilson} R.~W.,  2015,
  \mn@doi [\mnras] {10.1093/mnras/stv1400}, \href
  {http://adsabs.harvard.edu/abs/2015MNRAS.452.1707O} {452, 1707}

\bibitem[\protect\citeauthoryear{{Pepper}, {Stanek}, {Pogge}, {Latham},
  {DePoy}, {Siverd}, {Poindexter}  \& {Sivakoff}}{{Pepper}
  et~al.}{2008}]{2008AJ...135..907P}
{Pepper} J.,  {Stanek} K.~Z.,  {Pogge} R.~W.,  {Latham} D.~W.,  {DePoy} D.~L.,
  {Siverd} R.,  {Poindexter} S.,   {Sivakoff} G.~R.,  2008, \mn@doi [\aj]
  {10.1088/0004-6256/135/3/907}, \href
  {http://adsabs.harvard.edu/abs/2008AJ...135..907P} {135, 907}

\bibitem[\protect\citeauthoryear{{Pont}, {Sing}, {Gibson}, {Aigrain}, {Henry}
  \& {Husnoo}}{{Pont} et~al.}{2013}]{2013MNRAS.432.2917P}
{Pont} F.,  {Sing} D.~K.,  {Gibson} N.~P.,  {Aigrain} S.,  {Henry} G.,
  {Husnoo} N.,  2013, \mn@doi [\mnras] {10.1093/mnras/stt651}, \href
  {http://adsabs.harvard.edu/abs/2013MNRAS.432.2917P} {432, 2917}

\bibitem[\protect\citeauthoryear{{Rackham} et~al.,}{{Rackham}
  et~al.}{2017}]{2017ApJ...834..151R}
{Rackham} B.,  et~al., 2017, \mn@doi [\apj] {10.3847/1538-4357/aa4f6c}, \href
  {http://adsabs.harvard.edu/abs/2017ApJ...834..151R} {834, 151}

\bibitem[\protect\citeauthoryear{{Redfield}, {Endl}, {Cochran}  \&
  {Koesterke}}{{Redfield} et~al.}{2008}]{2008ApJ...673L..87R}
{Redfield} S.,  {Endl} M.,  {Cochran} W.~D.,   {Koesterke} L.,  2008, \mn@doi
  [\apjl] {10.1086/527475}, \href
  {http://adsabs.harvard.edu/abs/2008ApJ...673L..87R} {673, L87}

\bibitem[\protect\citeauthoryear{{Seager} \& {Sasselov}}{{Seager} \&
  {Sasselov}}{2000}]{2000ApJ...537..916S}
{Seager} S.,  {Sasselov} D.~D.,  2000, \mn@doi [\apj] {10.1086/309088}, \href
  {http://adsabs.harvard.edu/abs/2000ApJ...537..916S} {537, 916}

\bibitem[\protect\citeauthoryear{{Sing} et~al.,}{{Sing}
  et~al.}{2011}]{2011MNRAS.416.1443S}
{Sing} D.~K.,  et~al., 2011, \mn@doi [\mnras]
  {10.1111/j.1365-2966.2011.19142.x}, \href
  {http://adsabs.harvard.edu/abs/2011MNRAS.416.1443S} {416, 1443}

\bibitem[\protect\citeauthoryear{{Sing} et~al.,}{{Sing}
  et~al.}{2012}]{2012MNRAS.426.1663S}
{Sing} D.~K.,  et~al., 2012, \mn@doi [\mnras]
  {10.1111/j.1365-2966.2012.21938.x}, \href
  {http://adsabs.harvard.edu/abs/2012MNRAS.426.1663S} {426, 1663}

\bibitem[\protect\citeauthoryear{{Sing} et~al.,}{{Sing}
  et~al.}{2013}]{2013MNRAS.436.2956S}
{Sing} D.~K.,  et~al., 2013, \mn@doi [\mnras] {10.1093/mnras/stt1782}, \href
  {http://adsabs.harvard.edu/abs/2013MNRAS.436.2956S} {436, 2956}

\bibitem[\protect\citeauthoryear{{Sing} et~al.,}{{Sing}
  et~al.}{2015}]{2015MNRAS.446.2428S}
{Sing} D.~K.,  et~al., 2015, \mn@doi [\mnras] {10.1093/mnras/stu2279}, \href
  {http://adsabs.harvard.edu/abs/2015MNRAS.446.2428S} {446, 2428}

\bibitem[\protect\citeauthoryear{{Sing} et~al.,}{{Sing}
  et~al.}{2016}]{2016Natur.529...59S}
{Sing} D.~K.,  et~al., 2016, \mn@doi [\nat] {10.1038/nature16068}, \href
  {http://adsabs.harvard.edu/abs/2016Natur.529...59S} {529, 59}

\bibitem[\protect\citeauthoryear{{Siverd} et~al.,}{{Siverd}
  et~al.}{2012}]{2012ApJ...761..123S}
{Siverd} R.~J.,  et~al., 2012, \mn@doi [\apj] {10.1088/0004-637X/761/2/123},
  \href {http://adsabs.harvard.edu/abs/2012ApJ...761..123S} {761, 123}

\bibitem[\protect\citeauthoryear{{Snellen}, {de Kok}, {de Mooij}  \&
  {Albrecht}}{{Snellen} et~al.}{2010}]{2010Natur.465.1049S}
{Snellen} I.~A.~G.,  {de Kok} R.~J.,  {de Mooij} E.~J.~W.,   {Albrecht} S.,
  2010, \mn@doi [\nat] {10.1038/nature09111}, \href
  {http://adsabs.harvard.edu/abs/2010Natur.465.1049S} {465, 1049}

\bibitem[\protect\citeauthoryear{{Southworth}}{{Southworth}}{2010}]{2010MNRAS.408.1689S}
{Southworth} J.,  2010, \mn@doi [\mnras] {10.1111/j.1365-2966.2010.17231.x},
  \href {http://adsabs.harvard.edu/abs/2010MNRAS.408.1689S} {408, 1689}

\bibitem[\protect\citeauthoryear{{Stassun}, {Collins}  \& {Gaudi}}{{Stassun}
  et~al.}{2017}]{2017AJ...153..136S}
{Stassun} K.~G.,  {Collins} K.~A.,   {Gaudi} B.~S.,  2017, \mn@doi [\aj]
  {10.3847/1538-3881/aa5df3}, \href
  {http://adsabs.harvard.edu/abs/2017AJ...153..136S} {153, 136}

\bibitem[\protect\citeauthoryear{{Stevenson}}{{Stevenson}}{2016}]{2016ApJ...817L..16S}
{Stevenson} K.~B.,  2016, \mn@doi [\apjl] {10.3847/2041-8205/817/2/L16}, \href
  {http://adsabs.harvard.edu/abs/2016ApJ...817L..16S} {817, L16}

\bibitem[\protect\citeauthoryear{{Tsiaras}, {Waldmann}, {Rocchetto}, {Varley},
  {Morello}, {Damiano}  \& {Tinetti}}{{Tsiaras}
  et~al.}{2016}]{2016ApJ...832..202T}
{Tsiaras} A.,  {Waldmann} I.~P.,  {Rocchetto} M.,  {Varley} R.,  {Morello} G.,
  {Damiano} M.,   {Tinetti} G.,  2016, \mn@doi [\apj]
  {10.3847/0004-637X/832/2/202}, \href
  {http://adsabs.harvard.edu/abs/2016ApJ...832..202T} {832, 202}

\bibitem[\protect\citeauthoryear{{Tsiaras} et~al.,}{{Tsiaras}
  et~al.}{2018}]{2018AJ...155..156T}
{Tsiaras} A.,  et~al., 2018, \mn@doi [\aj] {10.3847/1538-3881/aaaf75}, \href
  {http://adsabs.harvard.edu/abs/2018AJ...155..156T} {155, 156}

\bibitem[\protect\citeauthoryear{{Turner} et~al.,}{{Turner}
  et~al.}{2016}]{2016MNRAS.459..789T}
{Turner} J.~D.,  et~al., 2016, \mn@doi [\mnras] {10.1093/mnras/stw574}, \href
  {http://adsabs.harvard.edu/abs/2016MNRAS.459..789T} {459, 789}

\bibitem[\protect\citeauthoryear{{Turner} et~al.,}{{Turner}
  et~al.}{2017}]{2017MNRAS.472.3871T}
{Turner} J.~D.,  et~al., 2017, \mn@doi [\mnras] {10.1093/mnras/stx2221}, \href
  {http://adsabs.harvard.edu/abs/2017MNRAS.472.3871T} {472, 3871}

\bibitem[\protect\citeauthoryear{{Vidal-Madjar}, {Lecavelier des Etangs},
  {D{\'e}sert}, {Ballester}, {Ferlet}, {H{\'e}brard}  \&
  {Mayor}}{{Vidal-Madjar} et~al.}{2003}]{2003Natur.422..143V}
{Vidal-Madjar} A.,  {Lecavelier des Etangs} A.,  {D{\'e}sert} J.-M.,
  {Ballester} G.~E.,  {Ferlet} R.,  {H{\'e}brard} G.,   {Mayor} M.,  2003,
  \mn@doi [\nat] {10.1038/nature01448}, \href
  {http://adsabs.harvard.edu/abs/2003Natur.422..143V} {422, 143}

\bibitem[\protect\citeauthoryear{{Wakeford}, {Visscher}, {Lewis}, {Kataria},
  {Marley}, {Fortney}  \& {Mandell}}{{Wakeford}
  et~al.}{2017}]{2017MNRAS.464.4247W}
{Wakeford} H.~R.,  {Visscher} C.,  {Lewis} N.~K.,  {Kataria} T.,  {Marley}
  M.~S.,  {Fortney} J.~J.,   {Mandell} A.~M.,  2017, \mn@doi [\mnras]
  {10.1093/mnras/stw2639}, \href
  {http://adsabs.harvard.edu/abs/2017MNRAS.464.4247W} {464, 4247}

\bibitem[\protect\citeauthoryear{{Winn}}{{Winn}}{2010}]{2010arXiv1001.2010W}
{Winn} J.~N.,  2010, preprint, \href
  {http://adsabs.harvard.edu/abs/2010arXiv1001.2010W} {} (\mn@eprint {arXiv}
  {1001.2010})

\bibitem[\protect\citeauthoryear{{Wyttenbach}, {Ehrenreich}, {Lovis}, {Udry}
  \& {Pepe}}{{Wyttenbach} et~al.}{2015}]{2015AA...577A..62W}
{Wyttenbach} A.,  {Ehrenreich} D.,  {Lovis} C.,  {Udry} S.,   {Pepe} F.,  2015,
  \mn@doi [\aap] {10.1051/0004-6361/201525729}, \href
  {http://adsabs.harvard.edu/abs/2015A\%26A...577A..62W} {577, A62}

\bibitem[\protect\citeauthoryear{{Young}}{{Young}}{1967}]{1967AJ.....72..747Y}
{Young} A.~T.,  1967, \mn@doi [\aj] {10.1086/110303}, \href
  {http://adsabs.harvard.edu/abs/1967AJ.....72..747Y} {72, 747}

\bibitem[\protect\citeauthoryear{{de Mooij} et~al.,}{{de Mooij}
  et~al.}{2012}]{2012AA...538A..46D}
{de Mooij} E.~J.~W.,  et~al., 2012, \mn@doi [\aap]
  {10.1051/0004-6361/201117205}, \href
  {http://adsabs.harvard.edu/abs/2012A\%26A...538A..46D} {538, A46}

\makeatother
\end{thebibliography}
%To get the bib do a pdflatex ___, bibtex ____, pdflatex ____x2

%%%%%%%%%%%%%%%%%%%%%%%%%%%%%%%%%%%%%%%%%%%%%%%%%%
\begin{table}
	\centering
\caption{Observations of HD~189733}
	\label{tab:obstable}
	\begin{tabular}{lcrrrr}
\hline
Date & Filter & $N_{\text{exp}}$ & Exp & Cadence & Airmass\\ 
 (UTC) & & &(sec) & (min) &\\ 
\hline
2016 08 02 & $u'$ & 43 & 180 & 6.1 & 1.06--1.93\\ 
2016 08 02 & $g'$ & 43 & 20 & 6.1 & 1.06--1.91\\
2016 08 02 & $r'$ & 41 & 16 & 6.1 & 1.06--1.92\\
2016 08 02 & $i'$ & 40 & 20 & 6.1 & 1.06--1.98\\ 
2016 08 02 & $z'$ & 44 & 30 & 6.1 & 1.06--1.97\\ 
\\
2017 06 22 & $u'$ & 18 & 320 & 9.5 & 1.05--1.35\\ 
2017 06 22 & $g'$ & 18 & 20* & 9.5 & 1.05--1.39\\ 
2017 06 22 & $r'$ & 18 & 12 & 9.5 & 1.05--1.36\\ 
2017 06 22 & $i'$ & 18 & 16* & 9.5 & 1.05--1.32\\
2017 06 22 & $z'$ & 18 & 30* & 9.5 & 1.05--1.33\\ 
\\
2017 08 01 & $u'$ & 45 & 270 & 8.3 & 1.05--1.79\\ 
2017 08 01 & $g'$ & 61 & 20 & 8.3 & 1.05--1.54\\ 
2017 08 01 & $r'$ & 60 & 14* & 8.3 & 1.05--1.57\\  
2017 08 01 & $i'$ & 53 & 18  & 8.3 & 1.05--1.49\\
2017 08 01 & $z'$ & 54 & 30  & 8.3 & 1.05--1.46\\ 
\hline
\multicolumn{6}{l}{\textit{Notes:} * Denotes an exposure duration that changed} \\
\multicolumn{6}{l}{within the night as a response to total count fluctuations.} \\
\multicolumn{6}{l}{These changes were never more than 15\% total exposure} \\
\multicolumn{6}{l}{length and were typically two second corrections.}
	\end{tabular}
\end{table}

\begin{table*}
	\centering
\caption{Comparison Star Description}
	\label{tab:comps}
	\begin{tabular}{llrrrl}
\hline
Star & Reference Photometry & \multicolumn{1}{c}{V} & \multicolumn{1}{c}{B$-$V} & Angular Distance & Notes \\ 
 & Source & (mag) & (mag) &(arcsec) & \\ 
\hline
HD  189733 &\citet{2010MNRAS.403.1949K}& 7.65 & 0.93 & 0.0 & \textit{Target Host for reference} \\ 
HD  345464 &\citet{2000AA...355L..27H}& 8.93 & 0.57 & 520.34 &   \\ 
HD  345459 &\citet{2000AA...355L..27H}& 8.09 & 1.05 & 523.14 &   \\ 
HD  345585 &\citet{2000AA...355L..27H}& 9.47 & 1.10 & 1132.40 &   \\ 
HD  345442 &\citet{2000AA...355L..27H}& 8.37 & 0.35 & 1594.96 & \textit{excluded all filters 2016 Aug 2 -- not in observed field}  \\ 
& &  &  &  & \textit{excluded u' filter (2017 June 22, 2017 Aug 1) -- over-exposed} \\  
\hline
	\end{tabular}  
\end{table*}

\begin{table*}
	\centering
	\caption{Extracted Relative Photometry, Full Machine-Readable File Available in Electronic Version}
	\label{table:photometry_table}
	\begin{tabular}{lccccccc}
Filter & BJD$_{\text{TDB}}$ & Rel Flux & Rel Flux Err & Airmass & X(FITS) &  Y(FITS)  & Exposure (sec)\\
\hline
g' &   2457602.711642  & 1.001134    &   0.000068 &      1.11	& 2659.38 &   2102.92 & 30\\
g' &   2457602.719215  & 1.000741    &   0.000068 &      1.10	& 2659.15 &   2108.18 & 30\\
g' &   2457602.746492  & 0.999497    &   0.000085 &      1.06	& 2659.19 &   2116.88 & 20\\
g' &   2457602.752047  & 0.998831    &   0.000085 &      1.06	& 2659.64 &   2116.78 & 20\\
g' &   2457602.757259  & 1.000997    &   0.000085 &      1.06	& 2659.42 &   2119.75 & 20\\
g' &   2457602.762018  & 0.999588    &   0.000085 &      1.05	& 2659.57 &   2119.27 & 20\\
... & & & & & & & \\
\hline
\multicolumn{8}{l}{\textit{Notes:} BJD$_{\text{TDB}}$ is as calculated from each exposures header mid-point UTC time.} \\
\multicolumn{8}{l}{Rel Flux (Rel Flux Err) is HD~189733's AIJ derived relative flux (error).}\\
\multicolumn{8}{l}{Airmass is as taken from exposure headers.}\\
\multicolumn{8}{l}{X(FITS) and Y(FITS) are the (x,y) pixel center location of HD~189733 as computed by AIJ.} \\
	\end{tabular}
\end{table*}

\begin{table*}
	\centering
	\caption{Fixed Model Parameters and Literature Sources}
	\label{table:fixed_table}
	\begin{tabular}{llr}
\hline
Parameter & Source & Value \\
\hline
$a/R$ & \citep{2010MNRAS.408.1689S} & 8.984 \\
Orbital Period & \citep{2015MNRAS.450.3101B} & 2.2185752 days\\
eccentricity & \citep{2010MNRAS.408.1689S} & 0.00\\
inclination & \citep{2010MNRAS.408.1689S}  & $85\fdg78$\\
\hline
limb-darkening coefficients & &\\
\hline
$u_a$ & \citep{2011AA...529A..75C} & 1.082\\
$u_b$ & \citep{2011AA...529A..75C} & -0.2315\\
$g_a$ & \citep{2011AA...529A..75C} & 0.7917\\
$g_b$ & \citep{2011AA...529A..75C} & 0.0347\\
$r_a$ & \citep{2011AA...529A..75C} & 0.5625\\
$r_b$ & \citep{2011AA...529A..75C} & 0.1661\\
$i_a$ & \citep{2011AA...529A..75C} & 0.440\\
$i_b$ & \citep{2011AA...529A..75C} & 0.1937\\
$z_a$ & \citep{2011AA...529A..75C} & 0.2099\\
$z_b$ & \citep{2011AA...529A..75C} & 0.3604\\
\end{tabular}
\end{table*}

\begin{table*}
	\centering
	\caption{Derived Mid-transit Times}
	\label{table:midtransit_table}
	\begin{tabular}{ccc}
\hline
UTC 2016 Aug 02 date &  UTC 2017 June 22 date & UTC 2017 Aug 01 date \\
$2457602.86483 \pm  0.00013$ BJD$_{\text{TDB}}$ & $2457926.77447 \pm 0.00016$ BJD$_{\text{TDB}}$ & $2457966.708803 \pm 0.000098$ BJD$_{\text{TDB}}$\\
\end{tabular}
\end{table*}

\begin{table*}
	\centering
	\caption{Single Filter $R_{\text{p}}/R_{\text{s}}$}
	\label{tab:results}
	\begin{tabular}{lccccc}
\hline
Star Spot Model & $u$ & $g$ & $r$ & $i$ & $z$\\
\hline
uncorrected & $0.15719^{+0.00074}_{-0.00078}$ & $0.1566^{+0.0011}_{-0.0011}$ & $0.1571^{+0.0013}_{-0.0013}$ & $0.1547^{+0.0010}_{-0.0010}$  & $0.15415^{+0.00092}_{-0.00096}$ \\ 
$\delta=0.01$,T=4000~K correction & $0.15648^{+0.00074}_{-0.00077}$ & $0.1560^{+0.0011}_{-0.0011}$ & $0.1566^{+0.0013}_{-0.0013}$ & $0.15422^{+0.00099}_{-0.00099}$ & $0.15372^{+0.00092}_{-0.00096}$ \\ 
$\delta=0.043$,T=4250~K & $0.15433^{+0.00073}_{-0.00076}$ &  $0.1541^{+0.0011}_{-0.0011}$ & $0.1551^{+0.0013}_{-0.0012}$ & $0.15296^{+0.00098}_{-0.00098}$ & $0.15263^{+0.00091}_{-0.00095}$\\ 
	\end{tabular}
\end{table*}

\pagebreak
\begin{figure*}
	\includegraphics[width=\textwidth]{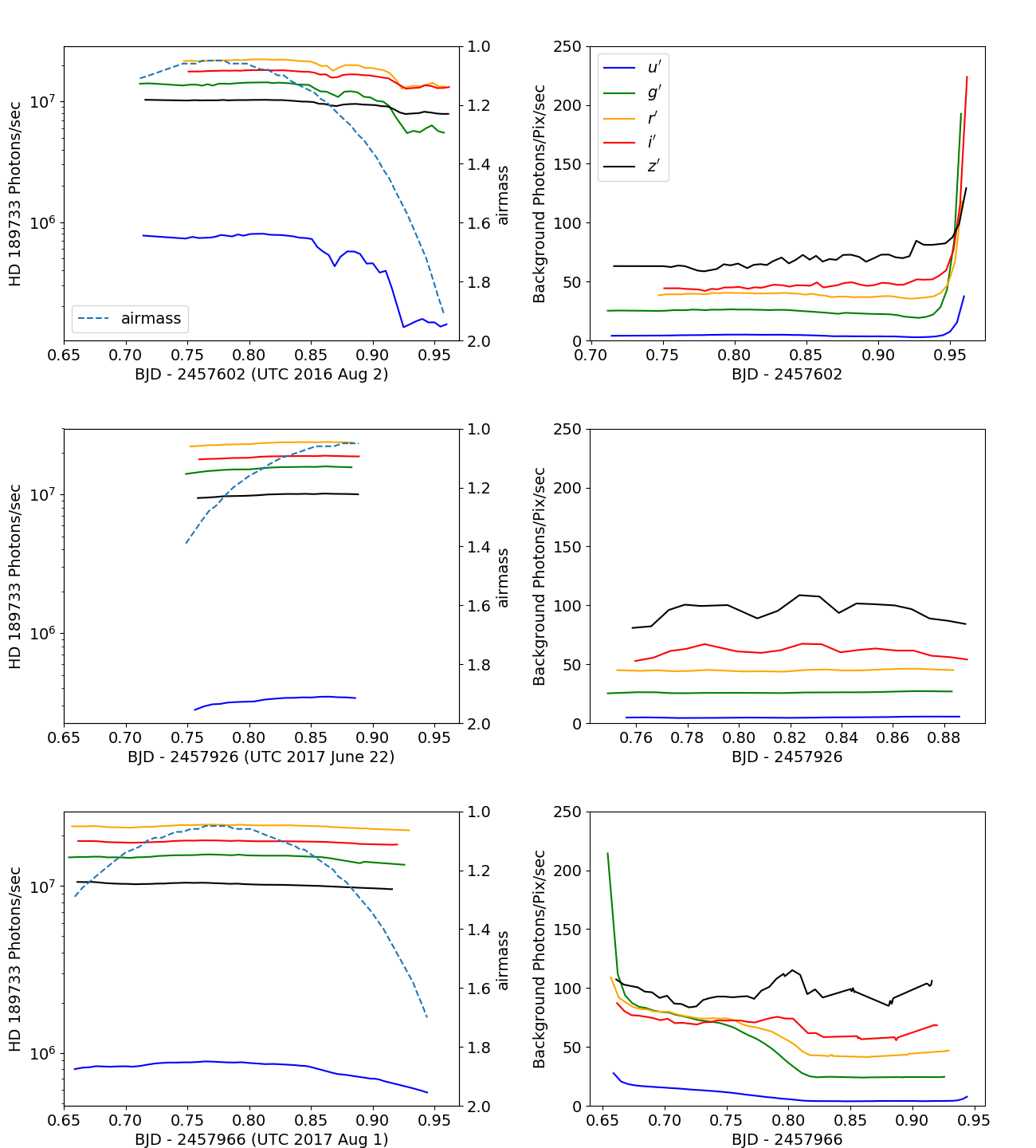}
    \caption{\textit{Left: (Top--Middle--Bottom)} 2016 Aug 02, 2017 June 22, 2017 Aug 01 raw target flux per second in $u'$ = blue to $z'$ = black with airmass plotted in light blue. \textit{Right:} Nightly corresponding calculated background flux per second per pixel color coded $u'$ = blue to $z'$ = black. The dramatic background flux rate rise at the end of the first night and fall at the beginning of the last night come from twilight.}
    \label{fig:weather_figure}
\end{figure*}

\pagebreak
\begin{figure*}
	\includegraphics[width=\textwidth]{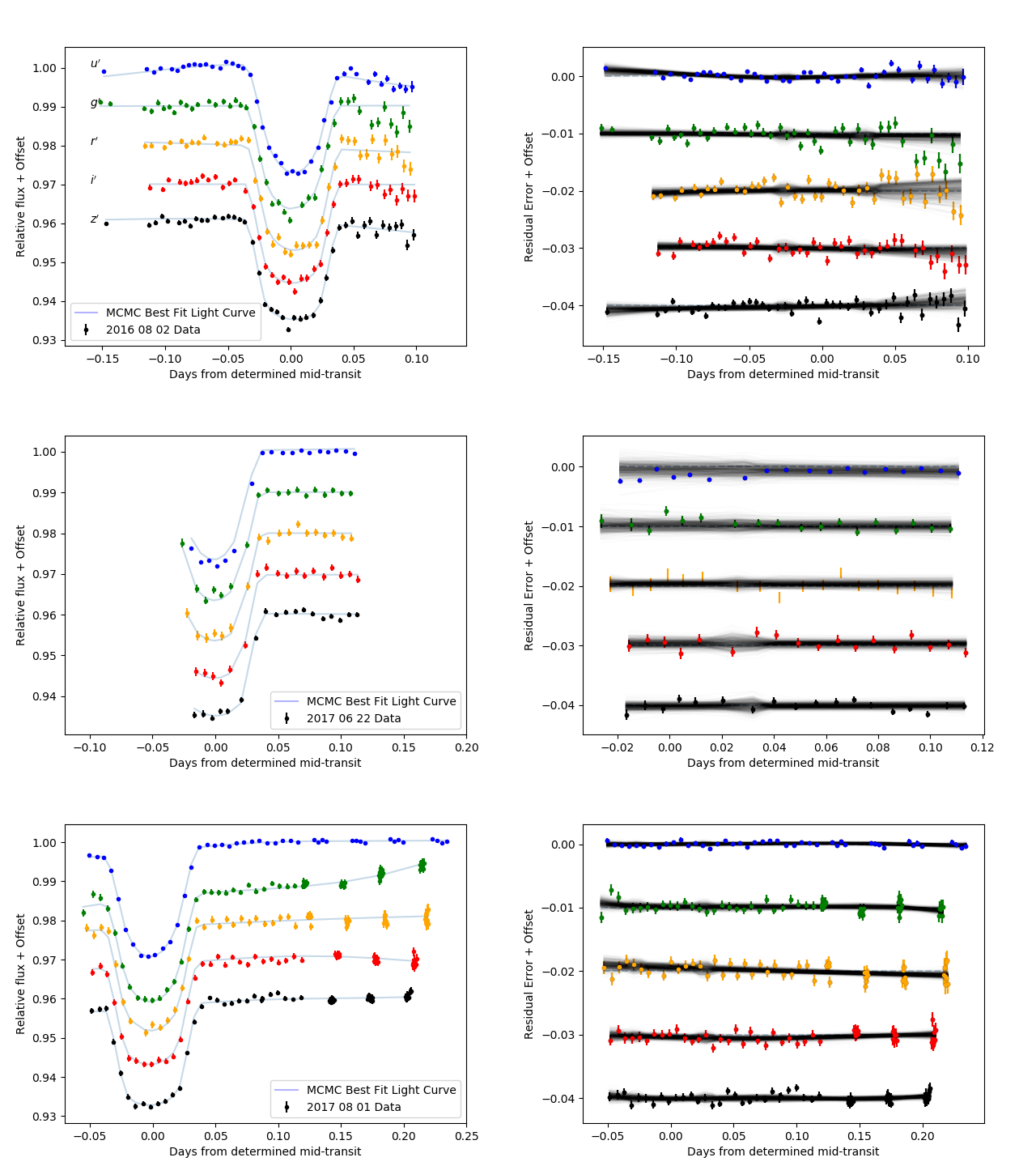}
    \caption{\textit{Left: (Top--Middle--Bottom)} 2016 Aug 02, 2017 June 22, 2017 Aug 01 HD~189733b transit observations in $u'$ = blue to $z'$ = black, with offsets for clarity, with light blue curves representing each data set's best fit model. \textit{Right:} Residuals of the data to the best fit models, with offset for clarity, with randomly sampled MCMC steps to indicate confidence.}
    \label{fig:multino_figure}
\end{figure*}

\pagebreak
\begin{figure}
	\includegraphics[width=\columnwidth]{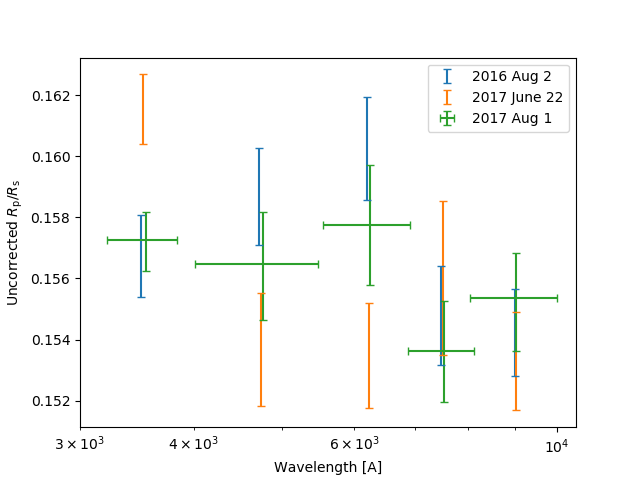}
    \caption{Each individual night's filter MCMC solved planet-to-star radius ratios (uncorrected for star spots) plotted against wavelength of observation color coded by night with small mid-point offsets for clarity. Horizontal errorbars indicate wavelength coverage.}
    \label{fig:soloradius_figure}
\end{figure}

\pagebreak
\begin{figure*}
	\includegraphics[width=\textwidth]{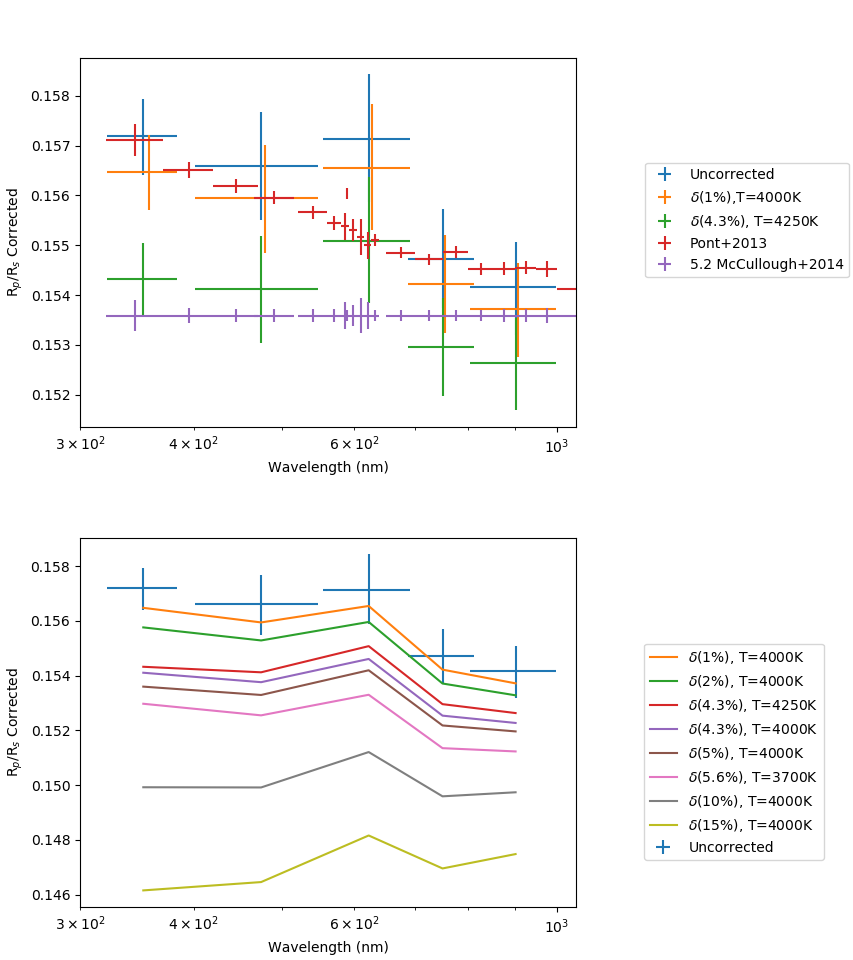}
    \caption{ \textit{Above:} Our multi-night, spot uncorrected $R_{\text{p}}/R_{\text{s}}$ versus wavelength (in blue) shown with the $\delta=0.01$, T=4000~K spot correction (in orange) and the $\delta=0.043$, T=4250~K spot correction (in green) with small mid-point offsets for clarity. Overplotted are optical \citet{2013MNRAS.432.2917P} results (in red) and the limiting case created in \citet{2014ApJ...791...55M} Section 5.2 of a grey atmosphere (in purple). All literature values have been offset by -0.001, to approximately correct for a difference in stellar limb darkening assumptions. We find agreement between our own potential corrections and the corresponding literature. \textit{Below:} Our multi-night, spot uncorrected $R_{\text{p}}/R_{\text{s}}$ versus wavelength (in blue) again, with a variety of potential spot coverage fractions and temperatures applied as color-coded. Each potential correction is visualized by a line connecting each filter's $R_{\text{p}}/R_{\text{s}}$ at mid-filter location. The atmospheric interpretation of our data, like literature sources, is highly dependent on the applied stellar spot correction.  }
    \label{fig:corrections_figure}
\end{figure*}

\begin{figure}
	\includegraphics[width=\columnwidth]{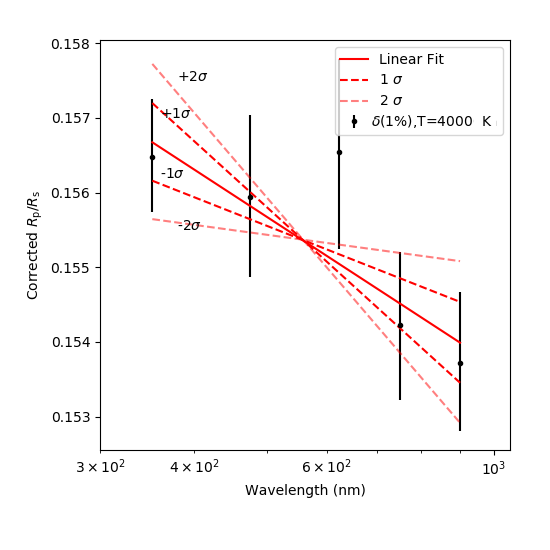}
    \caption{Our $R_{\text{p}}/R_{\text{s}}$ corrected for a spot temperature of 4000~K and coverage faction of $1\%$ in each filter (black points). 10,000-step, 100-walker MCMC linear fit is shown (solid red) with $1\sigma$ and $2\sigma$ slope uncertainties (dashed red).}
    \label{fig:mbfit_figure}
\end{figure}

% Don't change these lines
\bsp	% typesetting comment
\label{lastpage}
\end{document}